\documentclass[aps,prl,floatfix,showpacs,twocolumn,superscriptaddress]{revtex4}

\usepackage{graphics}
\usepackage{graphicx}
\usepackage{epsfig}
\usepackage{amssymb}
\usepackage{amsmath}
\usepackage{amstext}
\usepackage{bm}
\usepackage{times}

\renewcommand{\vec}[1]{\mathbf{#1}}
\newcommand{\expect}[1]{\ensuremath{\langle #1 \rangle}}
\newcommand{\ER}{\ensuremath{E_\text{R}}}
\newcommand{\intvecr}{\hspace{-0.05cm} \int \hspace{-0.7ex} d^3r \hspace{0.1cm} }
\newcommand{\Rb}{$^{87}$Rb }
\newcommand{\K}{$^{40}$K }
\newcommand{\B}{\ensuremath{\text{B}}}
\newcommand{\F}{\ensuremath{\text{F}}}
\newcommand{\BF}{\ensuremath{\text{BF}}}
\newcommand{\Vc}{\ensuremath{V_0^\text{c}}}

\newcommand{\AFFITP}{\affiliation{I. Institut f\"ur Theoretische Physik, Universit\"at Hamburg,
Jungiusstrasse 9, 20355 Hamburg, Germany} }
\newcommand{\AFFILP}{\affiliation{Institut f\"ur Laser-Physik, Universit\"at Hamburg,
Luruper Chaussee 149, 22761 Hamburg, Germany} }
\newcommand{\AFFMCUA}{\affiliation{Midlands Centre for Ultracold Atoms, School of Physics and Astronomy,
University of Birmingham, Edgbaston, Birmingham B15 2TT, United Kingdom} }

\begin{document}
\title{Self-Trapping of Bosons and Fermions in Optical Lattices}

\author{Dirk-S\"oren L\"uhmann}
\AFFITP
\author{Kai Bongs}
\AFFILP\AFFMCUA
\author{Klaus Sengstock}
\AFFILP
\author{Daniela Pfannkuche}
\AFFITP

\begin{abstract}

We  theoretically investigate  the enhanced localization of bosonic atoms by 
fermionic atoms 
in three-dimensional optical lattices and find a self-trapping  of the bosons 
for attractive boson-fermion interaction. 
Because of this  mutual interaction, the fermion orbitals are substantially squeezed,
which results in a strong deformation of the effective potential for bosons. 
This effect is enhanced by an increasing bosonic filling factor
leading to a large shift of the transition between the superfluid and the Mott-insulator phase.
We find a nonlinear dependency of the critical  potential depth on the boson-fermion  interaction strength.
The results, in general, demonstrate  the important
role of  higher Bloch bands for the physics of attractively interacting 
quantum gas mixtures in optical lattices  
and  are of direct relevance to recent experiments with $^{87}$Rb - $^{40}$K mixtures, 
where a large shift of the critical point has been found.

\end{abstract}

\date{31 July 2008}


\setlength{\textheight}{23.4cm}

~\\\maketitle

Ultracold multicomponent gases  offer new insight into  the atomic interspecies interaction.
In optical lattices, the interplay between interaction and tunneling is reflected
in the complex phase diagram of boson-fermion mixtures. 
The multitude of predicted phases 
have been studied theoretically using various approaches
\mbox{\cite{albus:023606,buchler130404,roth:021601,lewenstein:050401,cazalilla150403,cramer:190405,mathey:120404,pollet:190402, imambekov:021602,barillierpertuisel,rizzi:023621}}. 
In particular, the  phase separation between bosons and fermions,  the supersolid phase  
\cite{buchler130404,roth:021601},  and the  
pairing of bosons and fermions forming  phases of composite particles \cite{lewenstein:050401}
have been investigated.  
Mixtures of bosonic \Rb and fermionic \K atoms in optical lattices
were recently realized \cite{gunter:180402,ospelkaus:180403}
and  have drawn  much attention due to an unexpected large shift 
of the bosonic phase transition between the superfluid phase and the Mott insulator \cite{jaksch,zwerger}. 
The effect, which is substantial even for a small ratio of fermionic to bosonic atoms, 
was controversially discussed \cite{ospelkaus:180403,PhDSOspelkaus,pollet,cramer},
and a relatively small influence of the boson-fermion interaction 
on the transition was proposed.
In addition, the loss of coherence due to an adiabatic
heating in the optical lattice was addressed \cite{pollet,cramer}. 

In preference for a single-band Hubbard-type Hamiltonian, the influence of 
interaction induced orbital changes were widely neglected
in calculations performed for optical lattices.  
In this Letter, we show  by means of exact diagonalization  that the attractive interaction
between bosons and fermions 
causes substantially modified single-site densities. 
The nonlinearity of the interaction leads 
to a mutual trapping of  \Rb and \K atoms in the centers of the wells.   
For  high bosonic filling factors  the fermion orbitals,
i.e., the local atomic wave functions,   
are strongly squeezed, which leads to a considerable 
deformation of the  effective potential for bosons. 
The latter causes a large  shift of the  bosonic  Mott transition  towards shallower potentials
which   coincides with experimental results \cite{gunter:180402,ospelkaus:180403}.  
In this Letter, we show that the contribution of higher Bloch bands, usually neglected,
can be very important for the physics of ultracold quantum gases in optical lattices. 

The interaction between fully spin-polarized neutral atoms in the ultracold regime 
can be described by a contact potential $g\delta(\vec r -\vec r')$.
The strength of the repulsive interaction between \Rb atoms is given by $ g_\B=\frac{4\pi \hbar^2}{m_{\B}} a_{\B}$,
where $a_\B \approx 100a_0$ is the bosonic scattering length and $a_0$ the Bohr radius.
The interaction between \Rb and \K atoms far from a Feshbach resonance is attractive with the strength
$g_\BF=\frac{2\pi \hbar^2}{\mu} a_{\BF}$, where
$\mu=\frac{m_\B m_\F}{m_\B+m_\F}$ is the reduced mass and $a_{\BF} \approx -205a_0$ \cite{ferlaino:040702}.
The Hamiltonian  of a  Bose-Fermi mixture in an optical lattice 
is given by $\hat H= \hat H_{\B}+ \hat H_\F+ \hat H_{\BF}$, 
where $\hat H_\B$ describes the  system of interacting bosons,  
$\hat H_\F$ the system of noninteracting fermions,
and $\hat H_{\BF}$ the interaction between  bosons and fermions  \cite{albus:023606,buchler130404}.
Using the bosonic and fermionic field operators $\hat\psi_\B(\vec r)$ and $\hat\psi_\F(\vec r)$,
the three parts of the Hamiltonian can be written as 
\begin{eqnarray}
\hat H_\B&\hspace{-1ex}=& \hspace{-1ex} \intvecr \hat\psi_\B^\dagger(\vec r) \left[
     \frac{\hat {\vec p}^2}{2m_\B}+ V(\vec r)+ \frac{g_\B}{2} \hat\psi_\B^\dagger(\vec r) \hat\psi_\B(\vec r)
  \right]\hat\psi_\B(\vec r), \nonumber \\[1ex]
 \hat H_\F&\hspace{-1ex}=& \hspace{-1ex} \intvecr \hat\psi_\F^\dagger(\vec r) \left[
     \frac{\hat {\vec p}^2}{2m_\F}+ V(\vec r) \right] \hat\psi_\F(\vec r) ,\nonumber \\[1ex]
 \hat H_{\BF}&\hspace{-1ex}=& \hspace{-1ex} g_{\BF} \intvecr \hat\psi_\B^\dagger(\vec r) \hat\psi_\F^\dagger(\vec r)  \hat\psi_\F(\vec r) \hat\psi_\B(\vec r),
\label{H}	
\end{eqnarray}
with the periodic potential  $V(\vec r)$  of the optical lattice.

Numerically, we perform an exact diagonalization of the Hamiltonian (\ref{H}) in a many-particle basis \cite{Note0} 
that includes higher orbital states 
and is truncated at a sufficiently high energy (see \cite{luhmann:023620} for details). 
In our calculation,  
the bosonic and the fermionic subspace are diagonalized separately using a self-consistent interaction potential, 
which converges within  a  few cycles.
For a known fermionic density $\rho_\F(\vec r)=\expect{\hat \Psi_\F^\dagger \hat \Psi_\F}$ the effective 
Hamiltonian of the bosonic subsystem is given by
$\hat H_\B^\text{eff} (\rho_\F) =\hat H_\B+g_{\BF}  \intvecr  \rho_\F(\vec r)\, \hat\psi_\B^\dagger(\vec r) \hat\psi_\B(\vec r)$.
The latter term represents the interaction with the fermionic density and leads to a bosonic effective potential $V_\B^\text{eff}(\rho_\F)=V(\vec r) +g_{\BF} \rho_\F(\vec r)$.
Starting with the density of noninteracting fermions, 
we determine the boson density $\rho_\B(\vec r)=\expect{\Psi_\B^\dagger \Psi_\B}$ 
 by diagonalization of $\hat H_\B^\text{eff}$.  
Afterwards,  a new fermion density  can be  calculated  using the fermionic effective potential 
$V_\F^\text{eff}(\rho_\B)=V(\vec r)+g_{\BF} \rho_\B(\vec r)$.

\begin{figure}[b]
\includegraphics[width=0.85\linewidth]{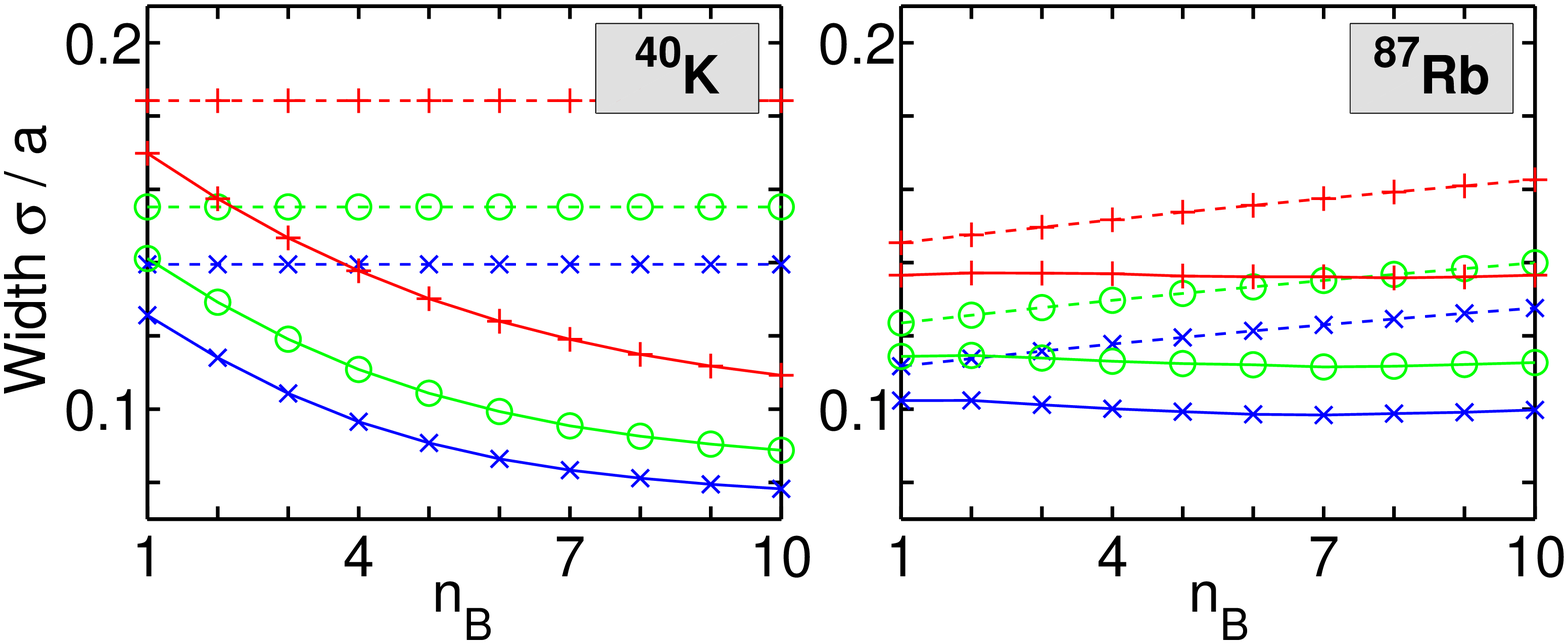}
\caption{(Color online)  The width of  the density profile  
for  one fermionic $^{40}$K atom (left) and $n_\B$ bosonic $^{87}$Rb atoms (right)  
for $V_0=8\ER$ ($+$), $V_0=14\ER$ ($\circ$), and $V_0=20\ER$ ($\times$).   
The dashed lines are obtained for vanishing boson-fermion interaction. 
Because of the mutual attraction, the densities are substantially squeezed.
The lattice constant $a=\pi/k$ is $515\text{ nm}$.  }
\label{FWHM}
\end{figure}

\begin{figure}[b]
\includegraphics[width=\linewidth]{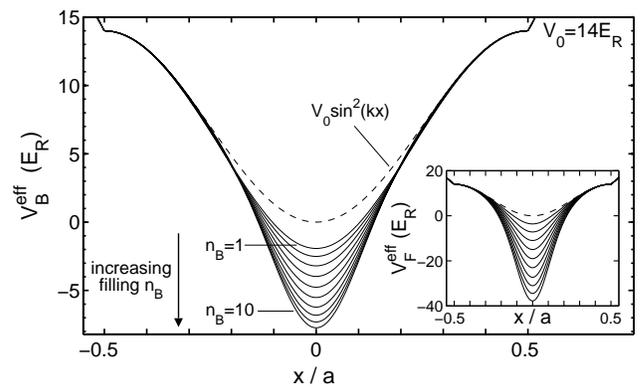}
\caption{The effective potential $V_\B^\text{eff}$ for $n_\B$ bosonic $^{87}$Rb atoms on one symmetric site 
created by one fermionic $^{40}$K atom.
The dashed line represents the unperturbed lattice potential $V(x)$ for vanishing boson-fermion interaction. 
The inset shows the  effective potential $V_\F^\text{eff}$ experienced by the fermion. }
\label{EffBPotential}
\end{figure}

We discuss the experimental situation, where one fermion ($n_\F=1$) and several bosons 
($1\leq n_\B \leq 10$) are present at each lattice   site. 
To study the effect caused by the mutual interaction,   
it is instructive  to restrict the particles to a single  lattice site.  
Hence, we perform the diagonalization for atoms in  a symmetric well with 
the shape $V_0[\sin^2(kx)+\sin^2(ky)+\sin^2(kz)]$ \cite{note1}. 
To illustrate the orbital changes, 
we fit Gaussians to the calculated densities for bosons and fermions 
$\rho_{\B/\F}(\vec r)=n_{\B/\F} (\sqrt{2\pi}\sigma_{\B/\F})^{-3} \exp(-r^2/2\sigma_{\B/\F}^2$)
\cite{note2}. 
The width of the Gaussians $\sigma_{\B/\F}$ 
as a function of the bosonic filling $n_\B$ 
is shown in Fig.~\ref{FWHM} for  the lattice depths  $V_0=8\ER$, $14\ER$, and $20\ER$,
where $\ER=\frac{\hbar^2 k^2}{2m_\B} $ is the recoil energy of \Rb atoms. 
Since \K atoms are much lighter than \Rb atoms, the resulting density profile for
a \K atom  on a single lattice site is much broader than for the \Rb atoms. 
For a  pure bosonic system  (dashed lines), we observe an increase
of the bosonic  width  with an increasing number of \Rb atoms, due to the boson-boson repulsion. 
The interaction with the single fermion leads to a compression of the boson density (solid lines), 
which becomes even narrower than the single-particle boson density.
In fact, the bosonic width decreases slightly with an increasing $n_\B$ 
and reaches a minimum at $n_\B=7-8$. 
This effective attractive behavior is surprising, since the attractive interaction
between bosons and fermions scales linearly with $n_\B$, whereas 
the repulsion of the bosons is proportional to $n_\B^2$. 
This effect is caused by the strong squeezing of the fermion density due 
to its effective potential $V_\F^\text{eff}=V(\vec r)+g_{\BF} \rho_\B(\vec r)$,
that is deepened linearly with the number of bosons $n_\B$ (inset of Fig.~\ref{EffBPotential}).  
 The increasing curvature of the fermionic effective potential, which equals
$\frac{\partial^2}{\partial x^2}V_\F^\text{eff}|_{\vec r=0} \approx 2 V_0 k^2 + \sqrt{2\pi}^{-3} |g_{\BF}|\, n_\B/\sigma_\B^5 $ 
using the Gaussian approach,
causes  the width of the fermion density to be similar to the bosonic one 
for $n_\B=3-4$ and even narrower for $n_\B>4$.  
In our calculation, the occupation of higher single-particle fermion orbitals,
due to the squeezing of the fermion density, is roughly  $10\%$ for $n_\B=4$ and
reaches $40\%$ for $n_\B=10$. 

The  effective potential $V_\B^\text{eff}(\rho_\F)$ 
experienced by the \Rb atoms is plotted in Fig.~\ref{EffBPotential} for $V_0=14\ER$.  
The curvature of the the bosonic effective potential
$\frac{\partial^2}{\partial x^2}V_\B^\text{eff}|_{\vec r=0} \approx 2V_0k^2 + \sqrt{2\pi}^{-3} |g_{\BF}| / \sigma_\F^5$
is strongly enhanced by the nonlinear dependence on the fermion width $\sigma_F$, 
which leads to a deepening of the bosonic effective potential with increasing $n_\B$.  
The resulting energy gain in $\expect{\hat H_{\BF}}$ overcompensates 
the repulsion energy of the bosons.  
Instead of a stronger repulsion, an increasing $n_\B$ causes 
a self-trapping of the bosons in the deepened center of the effective potential.
This effect is mediated by the squeezing of the fermion orbital due to the higher boson density.   
The compression of boson and fermion densities should be observable in the experiments
by imaging a  broadened  momentum distribution  or  by using  atomic clock shifts to
measure the peak density directly \cite{Campbell:313:649}.  

We now address the consequences of more than a single site but only one fermionic ``impurity."
We use the same diagonalization method as above that includes the deformation of orbitals
for small quasi-one-dimensional chains \cite{luhmann:023620}  
with $5-7$ sites, a bosonic filling factor $1\leq n_\B \leq 3$,
and $15\ER \leq V_0 \leq 20\ER$. 
The self-trapping causes a local deepening of the bosonic effective
potential for the fermion occupied site. 
The gain in energy, due to the  boson-fermion  interaction, leads to a binding of several
bosons to one fermion, which is a phenomenon resembling polaron physics \cite{cucchietti:210401,bruderer:011605}. 
Remarkably, we find that the fermionic impurity causes, for the above parameters, 
the localization of  six additional bosons  at  its site. 
In experimental  setups  an additional confining potential is used \cite{gunter:180402,ospelkaus:180403},
which establishes a finite atomic cloud,  so that   
bosons and fermions occupy only the central lattice sites.
Because of the self-trapping,  the bosons are attracted to sites
that are occupied by fermions.
If the fermionic filling $n_\F<1$ or the fermionic cloud is smaller than the bosonic one (as in Ref.~\cite{ospelkaus:180403}),
this leads to an increase of the bosonic filling factor in
mixtures in comparison with pure bosonic systems.   
In experiments, the predicted high bosonic site occupation will be
accompanied by high losses due to three-body recombinations. 

\begin{figure}
\includegraphics[width=\linewidth]{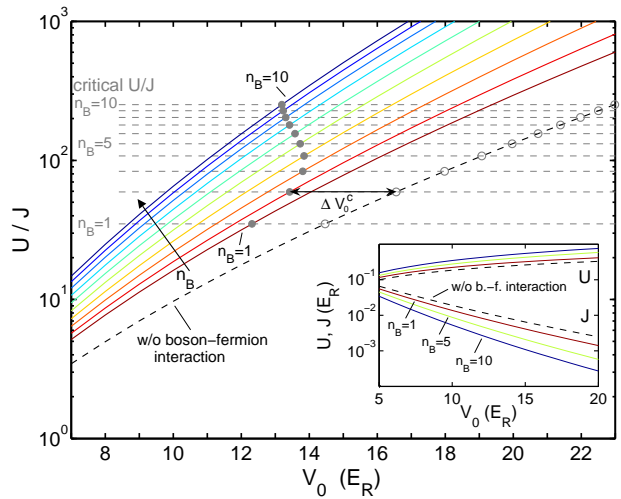} 
\caption{(Color online) Calculated ratio $U/J$ plotted against the lattice depth $V_0$ 
for filling factors $n_\B=1$ to $n_\B=10$ and $n_\F=1$ (solid lines). 
The dashed black line represents a pure bosonic system. 
The boson-fermion interaction causes a large shift of the critical potential depth $\Vc$.   
The inset  shows  $U$ and $J$ separately for the filling factors $n_\B=1$, $5$, $10$ and 
for vanishing interaction.}
\label{UJ}
\end{figure}

In the following, we discuss commensurately filled macroscopic cubic lattices and
the implications on the Mott insulator phase transition that result from the deformation of the effective
potential.  
In  optical lattices,  the bosonic  superfluid to Mott insulator transition is triggered by
the  boson-boson interaction strength relative to the hopping amplitude.
In the Bose-Hubbard framework \cite{jaksch,HubbardToolbox,zwerger},
this is described by the ratio between the on-site interaction $U$
and the hopping parameter $J$  that both depend on the depth of the lattice potential $V_0$.  
For boson-fermion mixtures, both $U$ and $J$ also depend explicitly 
on the filling factor.
However, to estimate how the the interaction with the fermions influences the phase transition,
we use an effective Bose-Hubbard model,  
where the lattice potential is  substituted by the bosonic effective potential  $V_\B^\text{eff}(n_\B)$
for a specific static filling factor $n_B$.
This approach 
leads to a renormalization of the parameters $U$ and $J$ 
in the effective bosonic system. 
The on-site interaction $U$ is obtained directly from $V_\B^\text{eff}(n_\B)$ 
and the hopping $J$ by band-structure calculations using a finite
periodic continuation of $V_\B^\text{eff}(n_\B)$ \cite{note4}. 
Qualitatively,  the bosonic effective potential,  which is shown in Fig.~\ref{EffBPotential}, 
reveals two important aspects:
First,  even  for a single boson the minimum is  deepened  substantially and decreases further with an increasing 
number of \Rb atoms; second, the shape of the effective potential deviates strongly
from the $V_0\sin^2(kx)$ potential, leading to a broader barrier between neighboring lattice sites.
Both effects lead to a reduced hopping between neighboring sites, whereas
the on-site interaction energy is mainly increased by the  stronger curvature of the effective  potential.
The decrease in $J$,   which is larger than the effect on $U$ (inset of Fig.~\ref{UJ}), 
causes narrower bands.

The ratio $U/J$  of the effective potential    $V_\B^\text{eff}(n_\B)$   is plotted in Fig.~\ref{UJ}  as a function of the lattice depth $V_0$,  where the 
dashed black line corresponds to a pure bosonic  system.  
The interacting system is represented by solid lines
which are split for different filling factors $n_\B$,
since  the deformation of the effective potential  grows with an increasing filling factor $n_\B$.
Even for  $n_\B=1$  the  self-trapping   
causes a large shift  of $U/J\,(V_0)$  towards higher values,  
which is further enhanced with an increasing  $n_\B$. 
In the following, we use the renormalized values of $U/J$ 
and  known  results for the critical point $(U/J)_\text{c}$  
to calculate the shift of   the critical potential depth $\Vc$. 
Following the mean field results, the  critical ratio  depends on the filling factor $n_\B$ and  obeys the 
relation $(U/J)_\text{c}=z [ 2n_\B+1+2\sqrt{n_\B(n_\B+1)} ]$,   
where $z=6$  for a cubic lattice \cite{krauth}. 
In Fig.~\ref{UJ}, the critical values for different filling factors  $n_\B$ are depicted  
as horizontal grey lines. 
The critical lattice  depths $\Vc(n_\B)$ for the  pure bosonic system 
are given by the intersections  of the dashed black line 
with the  horizontal lines (open circles). 
According to the relation above, the phase transition 
for a pure bosonic system shifts to deeper lattices for an increasing filling factor $n_\B$.
For boson-fermion mixtures, the intersections representing the
critical potential depths are  indicated by solid circles. 
In comparison with the pure bosonic system, 
the phase transition in mixtures is shifted substantially towards {\it shallower} lattices.

The  parameters  in our  calculations  have been chosen such that a comparison with
the experiment in Ref.~\cite{ospelkaus:180403} is possible.
The critical potential depth can be estimated from Fig.~\ref{UJ} 
as $12.3\ER$ for $n_\B=1$, increases to   $13.8\ER$  for $n_\B=4$,
and decreases slightly  for higher fillings.
Comparing  the boson-fermion mixture and the pure bosonic system  with the same bosonic filling,
the shift $\Delta \Vc$ of the phase transition is given by the difference 
of the respective critical lattice depths (see  arrow for $n_\B=2$).  
The expected shift increases substantially with the filling factor, e.g.,  
$\Delta \Vc=-2.2\ER$ for $n_\B=1$ and $\Delta \Vc=-5.2\ER$ for $n_\B=4$.  
In Refs.~\cite{gunter:180402,ospelkaus:180403}, pure bosonic systems are compared 
to Bose-Fermi mixtures with the same number of bosonic atoms, 
not accounting for the increase of the  local  bosonic filling factor
due to the self-trapping, as discussed above.   
In addition, the confinement in experiments causes the formation of shells with different filling factors $n_\B$.
Therefore, the reported shift in Ref.~\cite{ospelkaus:180403} of approximately $-5\ER$ 
is an average value of a system, where at the center of the lattice $n_\B>5$ and $n_\F=1$.
Accounting for the inhomogeneous filling,
we average our results for $n_\B\leq 5$ to $n_\B\leq 7$ 
leading to a shift between $-4.2\ER$ and $-5.1\ER$,
which is in good agreement with the experiment. 
Using atomic clock shifts \cite{Campbell:313:649}, 
the local filling factors   could be determined, which would allow a more accurate
comparison of experiment and theory. 

\begin{figure}[t]
\includegraphics[width=0.85\linewidth]{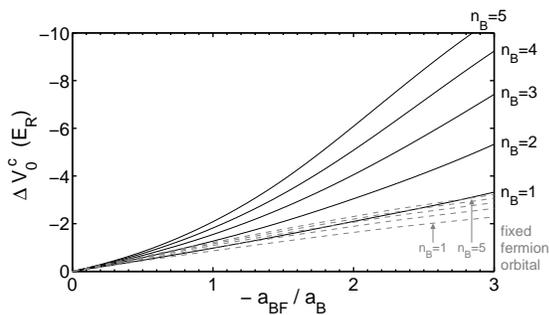} 
\caption{The shift of the superfluid to Mott insulator phase transition $\Delta \Vc$ 
in dependence on the scattering length  ratio $a_{\BF} / a_{\B}$ for 
the filling factors $n_\B=1$ to $n_\B=5$.
The dashed lines are obtained by keeping the fermion orbital fixed.}
\label{DeltaVc}
\end{figure}

Uniquely, experiments with ultracold atoms allow 
the precise tuning of the interaction strength by Feshbach resonances. 
Hence, the shift of the critical  potential depth  $\Delta \Vc$ can also be studied in
dependence on the scattering length  $a_{\BF}$  between bosons and fermions, while
the bosonic scattering length  $a_{\B}=100a_0$  is kept constant.
Such an experiment would allow a closer investigation of the mutual interaction and 
the effects due to orbital changes. 
In Fig.~\ref{DeltaVc}, we present the calculated shifts $\Delta \Vc$  
for bosonic filling factors $n_\B=1$ to $n_\B=5$,
where the solid lines are calculated allowing for orbital deformations and
the dashed lines by using a  rigid fermion orbital 
obtained by $V_\F^\text{eff}(\vec r)=V(\vec r)$.   
The latter  corresponds to a single-band approach for fermions,
leaving the orbital degrees of freedom only to the bosonic subsystem.
Because of a much weaker deformation of the  bosonic  effective potential, 
both $U$ and $J$ are less  affected  than in the previous discussion \cite{note5}. 
The scattering length $a_{\BF}$, which enters linearly in the Hamiltonian (\ref{H}), 
leads to an almost linear shift $\Delta \Vc$. 
In great contrast, the dependence  on $a_{\BF}$  for the self-consistent
calculation (solid lines), which fully includes the orbital degrees of freedom,
is superlinear due to the  self-trapping. 
Thus, the mutual deformation of the effective  potential  is enhanced
by a larger scattering length as well as by a higher bosonic filling factor
as discussed above.
Therefore, the  assumption of a fixed fermion orbital  is only suited for
weak interaction and low bosonic filling.  

For {\it repulsive} interaction between bosons and fermions ($a_{\BF}>0$)
the shift $\Delta \Vc$ becomes positive but remains rather small
with a sublinear dependence on $a_{\BF}$. 
In this region, however, the separation of bosons and fermions plays a role, which
is not accounted for in our calculations. 

In conclusion,  we have shown that orbital 
changes in attractive Bose-Fermi mixtures (\mbox{$^{87}$Rb -} $^{40}$K)  
are nonnegligible as they  
lead to a substantial deformation of the effective potential and
a squeezing of the effective orbitals.
The results are therefore of fundamental importance for
quantum gas mixtures in optical lattices. 
We found  a self-trapping behavior  of the bosons
in their effective potential mediated by the interaction with
the fermions. 
Using a model with  effective Bose-Hubbard  parameters $U$ and $J$, 
the expected shift of the  critical potential depth 
separating the superfluid phase from the Mott insulator
was estimated. 
Our results reveal a strong dependence of the shift on the  boson-fermion  interaction strength
and the bosonic filling factor,   
which could explain a large shift as reported in Refs.~\cite{gunter:180402,ospelkaus:180403}, 
and are applicable to future experiments using Feshbach resonances to tune the
interspecies interaction.
Theoretically, the full inclusion of orbital changes is a challenge for
the efficient calculation of lattice systems.  

We thank F. Deuretzbacher for helpful discussions 
and K. Patton for reading the manuscript.  
K. B. thanks EPSRC for financial support in Grant No.~EP/E036473/1.


\begin{thebibliography}{30}
\bibitem[{not({\natexlab{a}})}]{note1}
\bibinfo{note}{A smooth boundary is attached to the $\sin^2$-shaped single-site
  potential (see \cite{luhmann:023620}). A harmonic approach as in
  Refs.~\cite{ospelkaus:180403,PhDSOspelkaus} leads to noticeably narrower \K
  densities.}
\bibitem[{not({\natexlab{b}})}]{note2}
\bibinfo{note}{The deviations from the correct shape are relatively small using
  $\sigma_{\B/\F}=(n_{\B/\F}/\rho_{\B/\F}(0))^{1/3}/\sqrt{2\pi}$.}
\bibitem[{not({\natexlab{c}})}]{note4}
\bibinfo{note}{The parameter $J$ is determined as $1/4$ of the width of the
  lowest Bloch band of a chain with 30 sites \cite{HubbardToolbox}, and $U$ is
  the interaction of two particles $\expect{\frac{g_{\B}}{2} \intvecr
  \hat\psi_\B^{\dagger 2}(\vec r) \hat\psi_\B^2(\vec r) }$.}
\bibitem[{not({\natexlab{d}})}]{note5}
\bibinfo{note}{For a rigid fermion orbital, $U/J$ is independent of $n_\B$, but
  the difference from the pure bosonic system increases with the lattice depth.
  Thus, $\Delta \Vc$ increases slightly with $n_\B$.}
\end{thebibliography}
\end{document}